\documentclass[preprint,showpacs,prl]{revtex4}
\usepackage{bm}
\usepackage{epsfig}
\newcommand{\nix}[1]{}

\begin{document}

\title{
Magneto-Gyrotropic Photogalvanic Effect
\\in Semiconductor Quantum Wells }
\author{ V.V.~Bel'kov$^{1,2}$, S.D.~Ganichev$^{1,2}$, Petra~Schneider$^1$,
S.~Giglberger$^1$, E.L.~Ivchenko$^2$, S.A.~Tarasenko$^2$,
W.~Wegscheider$^1$, D.~Weiss$^1$, W.~Prettl$^1$}
\affiliation{$^1$Fakult\"{a}t Physik, University of Regensburg,
93040, Regensburg, Germany}
\affiliation{$^2$A.F.~Ioffe Physico-Technical Institute, Russian
Academy of Sciences, 194021 St.~Petersburg, Russia}


\begin{abstract}

We investigate both experimentally and theoretically, the
magneto-gyrotropic photogalvanic effect in zinc-blende based
quantum wells with $C_{2v}$ point-group symmetry using optical
excitation in the terahertz frequency range. The investigated
frequencies cause intra-subband but no inter-band and
inter-subband transitions. While at normal incidence the
photocurrent vanishes at zero magnetic field, it is shown that an
in-plane magnetic field generates photocurrents both for polarized
and unpolarized excitation. In general the spin-galvanic effect,
caused by circularly polarized light, and the magneto-gyrotropic
effect, caused by unpolarized excitation, is superimposed. It is
shown that in the case of two specific geometries both effects are
separable.

\end{abstract}

\maketitle

\section{Introduction}

The photogalvanic effect (PGE), predicted independently by
\cite{cpge,cpge1} for bulk semiconductors, is characteristic for
gyrotropic materials and was recently intensively studied, both
theoretically and experimentally, in zinc-blende and
diamond-lattice quantum well (QW) structures \cite{ufn,jphys}. In
such systems a photocurrent flows under illumination with
circularly polarized light which changes its direction if the
helicity of the circular polarization is reversed.
However, in the presence of a magnetic field a photocurrent can
flow even if the light is unpolarized~\cite{cpge,bulli}. This will
be denote as magneto-gyrotropic PGE below. The effect is due to
the fact that the gyrotropic point-group symmetry makes no
difference between components of polar and axial vectors and
therefore feature currents $j \propto I B$ with $I$ the light
intensity and $B$ the applied magnetic field. Here, the
proportionality constant is an invariant.

The magneto-gyrotropic PGE occurring under linearly polarized
irradiation in an applied magnetic field has been studied
theoretically in bulk crystals and
nanostructures~\cite{bulli,magarill,gorbats,kibis,spivak}.
Experimentally the effect was observed in QW
structures~\cite{emelya,moscow,kucher}, because
  the point groups
$D_{2d}$ and $C_{2v}$ of (001)-grown symmetrical and asymmetrical
QWs belong to gyrotropic classes. So far the magneto-gyrotropic
photocurrent has been observed for direct optical transitions. A
photocurrent caused by transitions between spin branches of the
electron subband splitted due to the Rashba effect has been
measured in a GaSb/InAs single QW structure (wavelength $\lambda$
= 385 $\mu$m or $\hbar \omega$ = 3.2 meV)~\cite{emelya}. A
magnetic field induced photocurrent due to direct inter-band
transitions  has also been observed in GaAs/AlGaAs QWs in the
spectral range 0.7 $\mu$m $< \lambda < 1.4$ $\mu$m and was
attributed to the magnetic field induced spin-independent
asymmetry of the electron dispersion in an asymmetric
heterostructure~\cite{gorbats,moscow,kucher}. Here we report on
the observation of the magneto-gyrotropic photocurrent in
$n$-doped InAs QWs under  intra-subband absorption ({\em indirect
Drude-like transitions}) of linearly and circularly polarized
far-infrared radiation ($\lambda$~=~148~$\mu$m and
$\lambda$~=~90~$\mu$m). The experimental results show that, in the
samples under study, the magneto-gyrotropic photocurrent has a
spin-related nature.

\section{Methods}

\begin{figure}
\centerline{\epsfysize 60mm \epsfbox{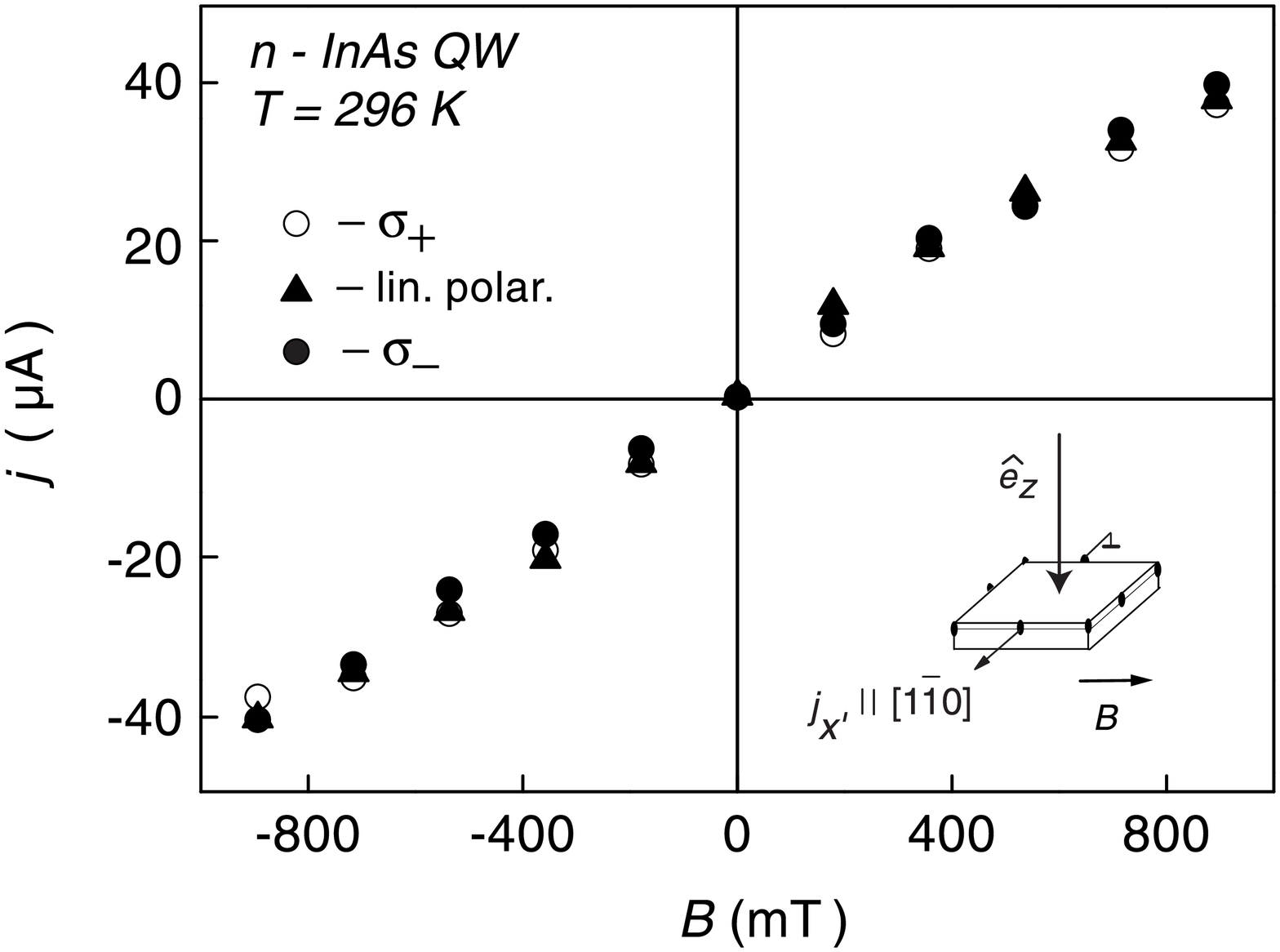}}
\caption{Magnetic-field dependence of the  photocurrent  measured
in InAs (001)-grown QWs at room temperature with magnetic field
$\bm B$ parallel to [110] direction. Optical excitation of 10~kW
power at normal incidence was applied at wavelength $\lambda =
148\:\mu$m with {\em linear}, right-handed {\em circular}
($\sigma_+$), and left-handed {\em circular} ($\sigma_-$)
polarization. The current is measured {\em normal} to $\bm B$. The
inset shows the geometry of the experiment. } \label{fig1}
\end{figure}

The experiments are  carried  out on (001)-oriented $n$-type
InAs/Al$_{0.3}$Ga$_{0.7}$Sb heterostructures having $C_{2v}$ point
symmetry. Single QWs of 15~nm width with free carrier densities of
about $1.3\cdot10^{12}$~cm$^{-2}$ and mobility $\approx
2\cdot10^4$~cm$^2$/(Vs) (data are obtained at room temperature)
were grown by molecular-beam epitaxy. Several samples of the same
batch were investigated at room temperature yielding the same
results. The samples have two pairs of  ohmic contacts at the
corners corresponding to the $\left< 100 \right>$ directions, $x
\parallel [100]$ and $y \parallel [010]$, and two additional pairs
of contacts centered along opposite sample edges with connecting
lines along $ x^\prime \parallel [1\bar{1}0]$ and $y^\prime
\parallel [110]$ (see insets in Figs.~\ref{fig1}\,-\,\ref{fig4}).
An external magnetic field $B$ up to $1$\,T was
applied parallel to the interface plane.

A pulsed optically pumped far-infrared laser was used for optical
excitation~\cite{PhysicaB99tun}. With  NH$_3$ as active gas 40~ns
pulses with about 10~kW power  have been obtained at the
wavelengths 148~$\mu$m and 90~$\mu$m. The far-infrared radiation
induces free carrier absorption (Drude-like)
in the lowest conduction subband $e$1 because the photon energies
are smaller than the subband separation and much larger than the
$\bm k$-linear
 spin splitting. The samples were irradiated along
the growth direction by linearly or circularly polarized
radiation. In all experiments the electric field vector of
linearly polarized radiation was oriented perpendicularly to the
magnetic field direction. Circular polarization was obtained by
using a crystalline quartz $\lambda/4$ plate. The helicity
$P_{circ}$ of the incident light was varied according to $P_{circ}
= \sin{2 \varphi}$, where $\varphi$ is the angle between the
initial plane of  linear polarization and the optical axis of the
$\lambda/4$ plate. $P_{circ}$ is equal to $ -1$ for left-handed
circularly polarized light $\sigma_-$ and $ +1$ for right-handed
polarization $\sigma_+$. The photocurrent $\bm j$ was measured at
room temperature in unbiased structures via the voltage drop
across a 50~$\Omega$ load resistor in a closed circuit
configuration.

\section{Experimental results}

\begin{figure}
\centerline{\epsfysize 60mm \epsfbox{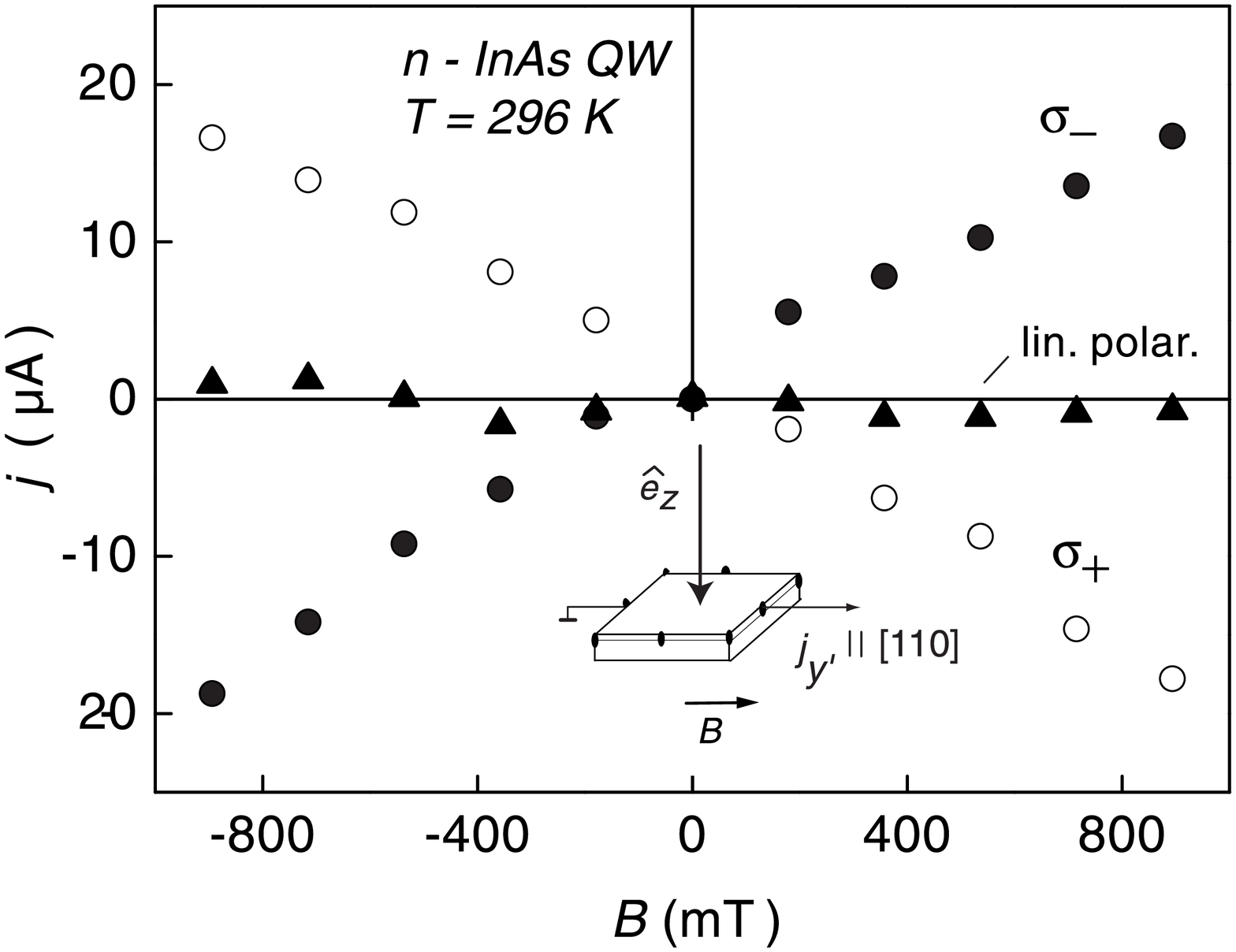}}
\caption{Magnetic-field dependence of the  photocurrent measured
with magnetic field $\bm B$ parallel to [110] direction. Optical
excitation at normal incidence was applied at wavelength $\lambda
= 148\:\mu$m with {\em linear}, right-handed {\em circular}
($\sigma_+$), and left-handed {\em circular} ($\sigma_-$)
polarization. The current is measured {\em parallel} to $\bm B$.
The inset shows the geometry of the experiment.} \label{fig2}
\end{figure}

\begin{figure}
\centerline{\epsfysize 60mm \epsfbox{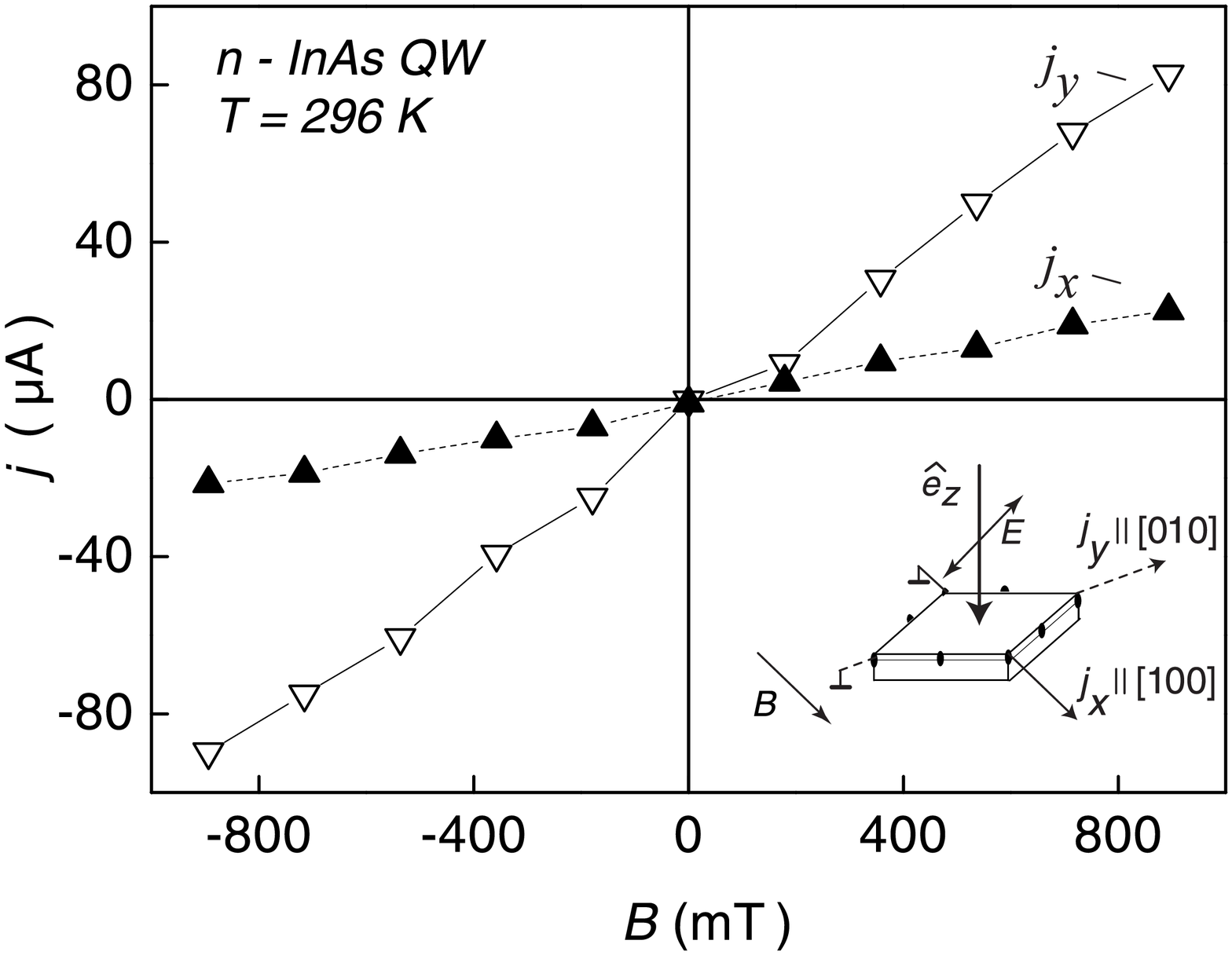}}
\caption{Magnetic-field dependence of  the photocurrent measured
with magnetic field $\bm B$ parallel to [100] direction. Optical
excitation at normal incidence was applied at wavelength $\lambda
= 148\:\mu$m with {\em linear} polarization. The current is
measured {\em parallel} ($j_x$) and {\em normal} ($j_y$) to $\bm
B$. The inset shows the geometry of the experiment.} \label{fig3}
\end{figure}

By irradiating the QWs with normal incident {\em linearly}
polarized light (see inset in Fig.~\ref{fig1}) a fast photocurrent
signal has been observed after applying an in-plane magnetic
field. The signal follows the temporal structure of the laser
pulse intensity, and the polarity of the current changes upon
reversal of the applied magnetic field. The results obtained for
$\lambda~=~90~\mu$m and for $\lambda~=~148~\mu$m are qualitatively
the same and differ only by a factor. Therefore below we present
only  data obtained for radiation with $\lambda~=~148~\mu$m. For
$\bm{B}$ aligned along a $\left< 110 \right>$ axis only a
transverse effect, i.e. current flow in the direction
perpendicular to the applied magnetic field, was detected (see
triangles in Figs.~\ref{fig1} and ~\ref{fig2}). For another
experimental configuration, $\bm{B}
\parallel \left< 100 \right>$, both longitudinal and transverse
currents were observed, depicted  in Fig.~\ref{fig3}. In the
absence of a magnetic field the signals vanish for all directions.
This is the case for {\em linear} as well as {\em circular} polarization.

A magnetic field induced current has also been observed upon
excitation with {\em circularly} polarized radiation. However, the
orientation of the current with respect to the direction of the magnetic field
and crystallographic axes is completely different. While  for $\bm B \parallel
\left< 110 \right> $
absorption of {\em linearly} polarized light  results in a photocurrent perpendicular to
the magnetic field direction only, the excitation by {\em
circularly} polarized radiation yields a current in both
directions, normal and parallel to the magnetic field
(Figs.~\ref{fig1} and \ref{fig2}). The current component normal to
the magnetic field, applied along $y^\prime
\parallel [110]$, is independent of the radiation helicity and
coincides with that induced by {\em linearly} polarized radiation
(Fig.~\ref{fig1}), say $j_{x^{\prime}} \propto I B_{y^{\prime}}$
where $I$ is the intensity of the radiation. This observation
indicates that the origin of this current is the same  for both
{\em linearly} and {\em circularly} polarized radiation. In
contrast, the current along the magnetic field direction changes its sign
upon changing the helicity from right- to left-handed and vanishes
for {\em linearly} polarized radiation: $j_{y^{\prime}} \propto I
B_{y^{\prime}} P_{circ}$ (see Fig.~\ref{fig2}). This current
has been observed previously~\cite{Nature02} and is caused by the
spin-galvanic effect due to optical orientation of carriers and
the Larmor precession of electron spins.

\begin{figure}
\centerline{\epsfysize 60mm \epsfbox{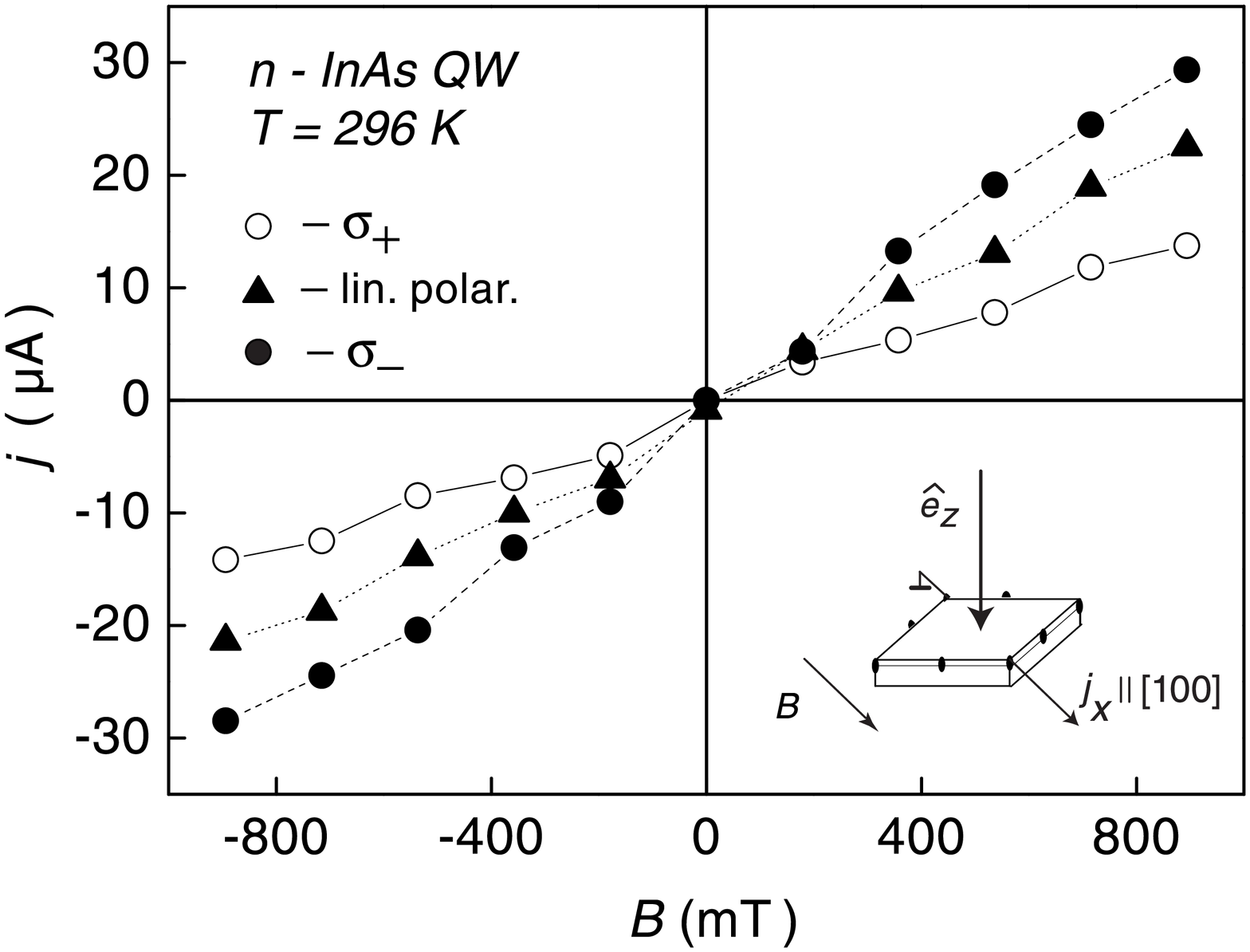}}
\caption{Magnetic-field dependence of  the photocurrent measured
with magnetic field $\bm B$ parallel to [100] direction. Optical
excitation at normal incidence was applied at wavelength $\lambda
= 148\:\mu$m with {\em linear}, right-handed {\em circular}
($\sigma_+$), and left-handed {\em circular} ($\sigma_-$)
polarization. The current is measured {\em parallel} to $\bm B$.
The inset shows the geometry of the experiment.} \label{fig4}
\end{figure}

If  the magnetic field is applied along one of the cubic axes, say ${\bm B} \parallel x$, the
current generated by  irradiation with helicity $P_{circ}$ may be described by the empirical formula
\begin{equation} \label{1}
j_i = (a_i + b_i  P_{circ}) B_x I\hspace{5 mm} (i=x,y)
\end{equation}
and can be detected for both directions $x \parallel [100]$ and
$y \parallel [010]$ (see Fig.~\ref{fig4}). In this equation $a_i$ and $b_i$ are constants of the same order of magnitude. The
helicity dependent component of the current determined by $b_i$  is caused in  this geometry by the spin-galvanic
effect superimposed on a helicity independent contribution given by $a_i$.

\section{Phenomenology}

In the linear approximation in the magnetic field strength
$\bm{B}$ the magnetic field induced photogalvanic effect is
phenomenologically given by
\begin{equation} \label{phen0}
j_\alpha = \sum_{\beta\gamma\delta}
\phi_{\alpha\beta\gamma\delta}\:B_\beta\:\{E_\gamma
E^\star_\delta\} + \sum_{\beta\gamma}
\mu_{\alpha\beta\gamma}\:B_\beta \hat{e}_\gamma\:E^2_0\:P_{circ},
\end{equation}
where the first term on the right hand side gives the magneto-gyrotropic
photogalvanic effect and the second term is the magnetic field induced spin-galvanic effect.
In this equation,
$\bm{E}$ is the electric field of the radiation  with $\bm{E} =
E_0\:\bm{e}$, where $E_0$ is the real amplitude of the field and
$\bm{e}$ denotes  a complex polarization vector of unit length,
$|\bm{e}| = 1$. $\{E_\gamma E^\star_\delta\}$ is the  symmetrized
 product of
the electric field with its complex conjugate
\begin{equation} \label{sym}
\{E_\gamma  E^\star_\delta\}  = \frac{1}{2}\left(E_\gamma
E^\star_\delta + E_\delta  E^\star_\gamma\right).
\end{equation}
 The  vector
$\bm{\hat e}$ is the unit vector pointing in the direction of
light propagation. The fourth rank tensor $\bm{\phi}$ is symmetric
in the last two indices. It is a pseudo-tensor whereas $\bm{\mu}$
is a regular third rank tensor. While the second term on the right
hand side of Eq.~(\ref{phen0}) requires circularly polarized
radiation the first term may be non-zero even for unpolarized
radiation.

In the following we consider (001)-oriented quantum wells based on
III-V$-$compounds.
The symmetry of these quantum wells may belong to the point groups
$D_{2d}$ or $C_{2v}$ depending, respectively, on equivalence or
non-equivalence of the QW interfaces. The present experiments have
been carried out on $C_{2v}$ structures and, therefore, we will
discuss this symmetry only.

In order to describe the experiments it is convenient to use  both
Cartesian coordinate systems introduced above. In  the coordinate
system   $x^\prime \parallel [1 \bar{1} 0]$, $y^\prime
\parallel [110]$, $z\parallel [001]$, the  in-plane
axes lie in the crystallographic planes $(110)$ and $(1 \bar{1}
0)$ which are the mirror reflection planes containing the two-fold
axes of $C_{2v}$. In this coordinate system Eq.~(\ref{phen0}) can
be reduced to
\begin{eqnarray} \label{phen}
j_{x^\prime} = S_1  B_{y^\prime}I + S_2 B_{y^\prime}  \left(
|e_{x^\prime}|^2 - |e_{y^\prime}|^2 \right) I+ S_3 B_{x^\prime}
\left( e_{x^\prime} e^*_{y^\prime} + e_{y^\prime} e^*_{x^\prime}
\right)
I+ S_4 B_{x^\prime}   I P_{\rm circ}\:,\\
j_{y^\prime} = S'_1  B_{x^\prime}I + S'_2 B_{x^\prime}  \left(
|e_{x^\prime}|^2 - |e_{y^\prime}|^2 \right) I+ S'_3 B_{y^\prime}
\left( e_{x^\prime} e^*_{y^\prime} + e_{y^\prime} e^*_{x^\prime}
\right) I + S'_4  B_{y^\prime}I P_{\rm circ} \:.\nonumber
\end{eqnarray}
In the crystallographic directions $x \parallel [100], y
\parallel [010]$, Eqs.~(\ref{phen}) are rewritten as
\begin{equation} \label{phena}
j_x = S_1^+ B_x I + S_1^- B_y I - (S_2^+ B_x + S_2^- B_y)  \left(
e_x e_y^* + e_y e_x^* \right) I $$ $$ + ( S_3^+ B_x - S_3^- B_y )
\left( |e_x|^2 - |e_y|^2 \right) I  + ( S_4^+ B_x - S_4^- B_y ) I
P_{\rm circ}\:,
\end{equation}
$$
j_y = - S_1^- B_x I - S_1^+ B_y I + (S_2^- B_x + S_2^+ B_y) \left(
e_x e_y^* + e_y e_x^* \right)I $$ $$ + ( - S_3^- B_x + S_3^+ B_y )
\left( |e_x|^2 - |e_y|^2 \right) I  + ( - S_4^- B_x + S_4^+ B_y )
I P_{\rm circ}\:,
$$
where $S_l^{\pm} = (S_l \pm S'_l)/2$ ($l=1\dots4$). In these
equations we set for the intensity $I=E_0^2$. The parameters $S_1$
to $S_4$, $S^\prime_1$ to $S^\prime_4$ and $S^{\pm}_1$ to
$S^{\pm}_4$  expressed by the non-zero elements of the tensors
$\bm{\phi}$ and $\bm{\mu}$ allowed in  $C_{2v}$ symmetry  are
given in Table~\ref{t1} and Table~\ref{t1a}, respectively.
Equations~(\ref{phen}) show that the first terms on the right hand
side (parameters $S_1,S^\prime_1 $) yield a current in the plane
of the quantum well which is proportional to the intensity $I$ and
independent of the state of radiation polarization. This current
is induced even for unpolarized radiation. In the case of the
second and third terms
 linear polarization is needed in order to give a photocurrent.
 The photocurrent assumes a maximum for light polarized along
 ${x^\prime}$ or ${y^\prime}$
in the case of the second terms described by the parameters
$S_2,S^\prime_2$ or along the bisector of ${x^\prime}$ and
${y^\prime}$ for the third terms proportional to $S_3,S^\prime_3$.

Thus,  these terms do not contribute to the current if the
radiation is circularly polarized. The last terms in
Eqs.~(\ref{phen}) describe a current proportional to the helicity
of radiation being maximum for left- or right-handed circular
polarization and changing sign when the helicity $P_{circ}$ is
switched from  $+1$ to $-1$. At normal incidence of radiation in
all cases a magnetic field in the plane of the quantum well is
required to obtain a current. At zero magnetic field no current is
allowed at normal incidence of radiation  in the present $C_{2v}$
point group symmetry case and, in fact, it does not occur in
experiment. The relations between polarization and magnetic field
after Eqs.~(\ref{phen}) and Eqs.~(\ref{phena}) are summarized in
Table~\ref{t2}.

\begin{table}
\renewcommand{\arraystretch}{1.5}
\begin{tabular}{|r@{=}l|r@{=}l|}
\hline

$S_1$ &
$\frac{1}{2}(\phi_{{x^\prime}{y^\prime}{x^\prime}{x^\prime}}+
\phi_{{x^\prime}{y^\prime}{y^\prime}{y^\prime}})$ & $S^\prime_1$ &
$\frac{1}{2}(\phi_{{y^\prime}{x^\prime}{x^\prime}{x^\prime}}+
\phi_{{y^\prime}{x^\prime}{y^\prime}{y^\prime}})$ \\\hline

$S_2$ &
$\frac{1}{2}(\phi_{{x^\prime}{y^\prime}{x^\prime}{x^\prime}}-
\phi_{{x^\prime}{y^\prime}{y^\prime}{y^\prime}})$ & $S^\prime_2$ &
$\frac{1}{2}(\phi_{{y^\prime}{x^\prime}{x^\prime}{x^\prime}}-
\phi_{{y^\prime}{x^\prime}{y^\prime}{y^\prime}})$
\\\hline

$S_3$ &
$\phi_{{x^\prime}{x^\prime}{x^\prime}{y^\prime}}=\phi_{{x^\prime}{x^\prime}{y^\prime}{x^\prime}}$
& $S^\prime_3$ &
$\phi_{{y^\prime}{y^\prime}{x^\prime}{y^\prime}}=\phi_{{y^\prime}{y^\prime}{y^\prime}{x^\prime}}$\\\hline

$S_4$ & $\mu_{{x^\prime}{x^\prime}z}$ & $S^\prime_4$ &
$\mu_{{y^\prime}{y^\prime}z}$
\\\hline
\end{tabular}
\caption{Definition of the parameters $S_i$ and $S^\prime_i$
($i=1\dots4$) in Eqs.~(\ref{phen}) in terms of non-zero components
of the tensors $\bm{\phi}$ and $\bm{\mu}$ for coordinates
${x^\prime}
\parallel [1 \bar{1} 0]$, ${y^\prime} \parallel [110]$ and $z\parallel
[001]$. $C_{2v}$ symmetry and normal incidence of radiation along
$z$ is assumed.} \label{t1}
\end{table}

\begin{table}
\renewcommand{\arraystretch}{1.5}
\begin{tabular}{|r@{=}l|r@{=}l|}
\hline

$S^+_1$ & $\frac{1}{2}(\phi_{xxxx} +\phi_{xxyy})$ & $S^-_1$ & $\frac{1}{2}(\phi_{xyxx} +\phi_{xyyy})$ \\

& $-\frac{1}{2}(\phi_{yyxx} +\phi_{yyyy})$ & &
$-\frac{1}{2}(\phi_{yxxx} +\phi_{yxyy})$ \\\hline

$S^+_2$ & $\phi_{yyxy} = \phi_{yyyx} $&  $S^-_2$ & $\phi_{yxxy} = \phi_{yxyx}$ \\

& $-\phi_{xxxy} =- \phi_{xxyx} $&  & $-\phi_{xyxy} = -\phi_{xyyx}
$\\\hline

$S^+_3$ & $\frac{1}{2}(\phi_{xxxx} -\phi_{xxyy})$ & $S^-_3$ & $-\frac{1}{2}(\phi_{xyxx} -\phi_{xyyy})$ \\

& $\frac{1}{2}(\phi_{yyxx} +\phi_{yyyy})$ & &
$-\frac{1}{2}(\phi_{yxxx} -\phi_{yxyy})$ \\\hline

$S^+_4$ & $\mu_{xxz} = \mu_{yyz}$ & $S^-_4$ & $-\mu_{xyz} =
-\mu_{yxz}$\\\hline
\end{tabular}
\caption{Definition of the parameters $S_i$ and $S^\prime_i$
($i=1\dots4$) in Eqs.~(\ref{phen}) in terms of non-zero components
of the tensors $\bm{\phi}$ and $\bm{\mu}$ for coordinates
$x\parallel [100]$, $y\parallel [010]$ and $z\parallel [001]$.
$C_{2v}$ symmetry and normal incidence of radiation along $z$ is
assumed. } \label{t1a}
\end{table}

\section{Comparison to experimental results}

Now we analyze the phenomenological Eqs.~(\ref{phen}) with respect
to experiments. Three magnetic field orientations were applied,
namely, $\bm{B}\parallel y^\prime$, $\bm{B}\parallel x^\prime$,
and  $\bm{B}\parallel x$. Note that due to $C_{2v}$ point group
symmetry the geometry $\bm{B}~=~(B,0,0) \parallel x$ is same to
that for $\bm{B}~=~(0, -B,0) \parallel y$, however,
$\bm{B}\parallel x^\prime$ and $\bm{B}\parallel y^\prime$ are not
equivalent. We discuss the electric current parallel and normal to
the magnetic field in each case for three states of polarization:
{\em linear} as well as left-  and right-handed {\em circular}. In
the experiments with {\em linear} polarized radiation the electric
field vector is always oriented normally to $\bm{B}$.

\begin{itemize}

\item $\bm{B}\parallel y^\prime$ (Figs.~\ref{fig1} and
~\ref{fig2})\,:

In this configuration we find that  strength and sign of the
current measured in the direction normal to the magnetic field,
$x^\prime$,  is the same for all three states of polarization (for
{\em linear} polarization in this case $e_{x^\prime} \neq 0$,
$e_{y^\prime} = 0$) as  shown in Fig.~\ref{fig1}. This allows us
to conclude that the current in this geometry in our samples is
dominated by the first term in Eqs.~(\ref{phen}), $S_1\neq 0$,
while the second term  is negligible, $S_2 = 0$. Thus, the
observed magneto-gyrotropic effect is represented by a
polarization-independent magnetic field induced photocurrent. The
current along $y^\prime$ appears also, but for {\em circular}
polarization only. It has the same magnitude for right- and
left-handed {\em circular} polarization but changes sign if the
helicity is switched from $\sigma_+$ to $\sigma_-$ and vanishes
for  {\em linear} polarization (see Fig.~\ref{fig2}). This is due
to the spin-galvanic effect described by the last term in
Eqs.~(\ref{phen0}) and~(\ref{phen}).

\newpage

\item  $\bm{B}\parallel x^\prime$:

Qualitatively  we have   the same situation  as described above if
we exchange $x^\prime$ and $y^\prime$. The current densities,
however, differ quantatively. The magnitude of the photocurrent
parallel to the magnetic field and caused by {\em circularly}
polarized radiation is
smaller than that in the above case $\bm{B}\parallel y^\prime$.
The {\em helicity independent} and perpendicular to $\bm{B}$
current is, in contrast,  also different, but it exeeds
that in the previous configuration. This difference in
photocurrents is due to the fact that for $C_{2v}$ point symmetry
 the axes $[1\bar{1} 0]$ and $[110]$ are non-equivalent which
 is taken into account in Eqs.~(\ref{phen}) by introducing
 independent parameters
$S_i$ and $S^\prime_i$ ($i = 1\dots 4$).

\item $\bm{B}\parallel x$ or $\bm{B}\parallel y$ (Figs.~\ref{fig3}
and ~\ref{fig4})\,:

For an orientation of the magnetic field parallel to one of the
two in-plane $\left< 100 \right> $ axes of the material  we find
for {\em linearly} polarized radiation current components not only
perpendicular but, in contrast to the previous magnetic field
orientations, also parallel to $\bm{B}$ as displayed in
Fig.~\ref{fig3}. For {\em circularly} polarized radiation we
observe also the spin-galvanic current at contact pairs with
connecting lines parallel and normal to the magnetic field. The
spin-galvanic effect manifests itself by a difference of the
strength of the current for left- and right-handed {\em circular}
polarization. The spin-galvanic current is superimposed on the
helicity independent magneto-gyrotropic current (Fig.~\ref{fig4}).

\end{itemize}


    \begin{center}

 \begin{table}

        \begin{tabular}{|c|c|c|c|c|c|}
        \hline

        \multicolumn{2}{|l|}{} & \multicolumn{3}{c|}{magneto-gyrotropic effect} & \mbox{SGE} \\
        \cline{3-6}

        \multicolumn{2}{|l|}{\raisebox{3ex}[-3ex] {} } & \mbox{\bf 1$^{st}$ term} & \mbox{\bf 2$^{nd}$ term} & \mbox{\bf 3$^{rd}$ term} & \mbox{\bf 4$^{th}$ term} \\\hline\hline

        \rule[-3mm]{0mm}{9mm}

        & $j_{x^\prime}/I$ & 0 & 0 &$S_3 B_{x^\prime} \left(e_{x^\prime}e^*_{y^\prime}+e_{y^\prime}e^*_{x^\prime}\right)$ & $S_4 B_{x^\prime} P_{\rm{circ}}$ \\ \cline{2-6}
        \rule[-3mm]{0mm}{9mm}
        \raisebox{3ex}[-3ex] {$B\|x^\prime$} & $j_{y^\prime}/I$ & $S^\prime_1 B_{x^\prime}$ & $S^\prime_2 B_{x^\prime} \left(|e_{x^\prime}|^2-|e_{y^\prime}|^2\right)$ & 0 & 0 \\ \hline\hline

        \rule[-3mm]{0mm}{9mm}
        & $j_{x^\prime}/I $&$ S_1 B_{y^\prime}$ & $S_2 B_{y^\prime} \left(|e_{x^\prime}|^2-|e_{y^\prime}|^2\right)$  & 0 & 0 \\ \cline{2-6}
        \rule[-3mm]{0mm}{9mm}
        \raisebox{3ex}[-3ex] {$B\|{y^\prime}$} & $j_{y^\prime}/I $& 0 & 0 & $S^\prime_3 B_{y^\prime} \left(e_{x^\prime}e^*_{y^\prime}+e_{y^\prime}e^*_{x^\prime}\right) $ & $S^\prime_4 B_{y^\prime} P_{\rm{circ}} $  \\ \hline\hline

        \rule[-3mm]{0mm}{9mm}
        & $j_{x}/I$ & $S^+_1 B_x$ & $-S^+_2B_x \left(e_xe^*_y+e_ye^*_x\right) $& $S^+_3B_x \left(|e_x|^2-|e_y|^2\right)$ & $S^+_4 B_x  P_{\rm{circ}}$ \\ \cline{2-6}
        \rule[-3mm]{0mm}{9mm}
        \raisebox{3ex}[-3ex] {$B\|x$} &$ j_{y}/I $& $-S^-_1B_x$ &$ S^-_2B_x \left(e_x e^*_y+e_y e^*_x\right)$ & $-S^-_3B_x \left(|e_x|^2-|e_y|^2\right)$ & $-S^-_4 B_x P_{\rm{circ}} $ \\ \hline\hline

        \rule[-3mm]{0mm}{9mm}
        & $j_{x}/I$ & $S^-_1B_y$ & $-S^-_2B_y \left(e_xe^*_y+e_ye^*_x\right)$ & $-S^-_3B_y \left(|e_x|^2-|e_y|^2\right)$ &$ -S^-_4 B_y P_{\rm{circ}}$  \\ \cline{2-6}
        \rule[-3mm]{0mm}{9mm}
        \raisebox{3ex}[-3ex] {$B\|y$} & $j_{y}/I$ &$ -S^+_1B_y $& $S^+_2B_y \left(e_xe^*_y+e_ye^*_x\right)$ & $S^+_3B_y \left(|e_x|^2-|e_y|^2\right)$ & $S^+_4 B_y P_{\rm{circ}}$  \\ \hline

        \end{tabular}
\caption{Contribution of the different terms of
   Eqs.~(\ref{phen})
and Eqs.~(\ref{phena}) to the current at different magnetic field
orientations. The two left columns give the magnetic field and the
current, respectively.} \label{t2}
        \end{table}

\end{center}

\section{Microscopic model}

In general, the observed magneto-gyrotropic PGE descibed by the
parameters $S_1, S'_1$ in Eqs.~(\ref{phen}) can be attributed to
three possible mechanisms, two of them being spin-dependent and
one spin-independent or diamagnetic. Here, we briefly characterize
all, present qualitative estimations for each and discuss which
mechanism is most likely  responsible for the magneto-gyrotropic
photocurrent observed in the samples under study. In the first
mechanism the electric current appears due to an asymmetry of
spin-dependent {\it spin-conserving} energy relaxation processes
in a system of hot carriers heated by  free carrier
absorption~\cite{bulli}. Due to the Zeeman spin splitting of
electron spin states in the magnetic field, say ${\bm B} \parallel
x$, the difference in population of the spin branches $s_x = \pm
1/2$ is given by the ratio $g \mu_B B_x/\bar{E}$, where $g$ is the
electron $g$-factor, $\mu_B$ is the Bohr magneton, and $\bar{E}$
is the average electron energy. The light absorption leads to a
non-equilibrium symmetrical distribution of electrons within each
spin branch with the average energy different from that in
equilibrium. Taking into account the gyrotropic symmetry of the
QW, the matrix element for scattering ${\bm k}, s_x \to {\bm k}',
s_x$ of an electron by a phonon can be presented in the form
$M({\bm k}', s_x;{\bm k}, s_x) = P + s_x \sum_i R_{xi} (k'_i +
k_i)$, where $P, R_{xi}$ $(i=1,2)$ as functions of the involved
wave vectors have the same parities. Thus, the ratio of
antisymmetric to symmetric parts of the scattering probability
rate $W_{{\bm k}', s_x;{\bm k}, s_x}$ is given by $W^{(-)}/W^{(+)}
\sim \sum_i R_{xi} (k'_i + k_i)/P$. Since the antisymmetric
component of the electron distribution function decays within the
momentum relaxation time $\tau_p$, one can write for the
photocurrent
\begin{equation}
j_i \sim e N \frac{g \mu_B B_x}{\bar{E}} \left\langle W^{(+)}
\frac{R_{xi} (k'_i + k_i)}{P} \left[ \tau_p(k') \frac{\hbar
k'_i}{m^*} - \tau_p(k) \frac{\hbar k_i}{m^*} \right] \right\rangle
\sim e \tau_p \frac{R_{xi}}{\hbar P} \frac{g \mu_B B_x}{\bar{E}}
\eta I\:, \label{2}
\end{equation}
where $e$ is the electron charge, $m^*$ is the electron effective
mass, $\eta$ is the fraction of the energy flux absorbed in the QW
due to all possible indirect optical transitions, and the angle
brackets mean averaging over the electron energy distribution.
While deriving Eq.~(\ref{2}) we took into account the balance of
energy:
$$\sum_{{\bm k}' {\bm k}}(E_{\bm k} - E_{{\bm k}'}) W_{{\bm
k}',{\bm k}} = \eta I \ ,$$ where $E_{\bm k} = \hbar^2 k^2/2 m^*$.
The contributions to the current from electrons in the branches
$s_x = \pm 1/2$ are of opposite sign. However, they do not
compensate each other because of the magnetic-field induced
selective occupation of the branches.

The next mechanism to be considered is based on the asymmetry of
spin-flip processes and represents in fact the spin-galvanic
effect described by
\begin{equation}
j_i = Q_{ij} (S_j - S^{0}_j)\:,
\label{Q}
\end{equation}
where ${\bm S}$ is the average electron spin and ${\bm S}^{0}$ is
its equilibrium value governed by the ratio $g \mu_B B/\bar{E}$.
The observation of the spin-galvanic effect~\cite{Nature02} was
carried out under conditions where ${\bm S}^{0}$ was negligible.
Then the spin-galvanic current is generated as a result of the
spin relaxation which tends to depress the polarization. The same
is relevant to the helicity-dependent photocurrent described by
the coefficients $b_i$ in Eq.~(\ref{1}). A possible contribution
of the spin-galvanic effect to the polarization-independent
photocurrent can be interpreted as a light-induced depolarization
of the electronic spins followed by the spin relaxation tending to
restore the polarization to ${\bm S}^{0}$. As well as
in~\cite{Nature02} the asymmetry of spin-flip processes can be
related to the ${\bm k}$-linear terms in the electron effective
Hamiltonian, $\beta_{BIA} (\sigma_x k_x - \sigma_y k_y) +
\beta_{SIA} (\sigma_x k_y - \sigma_y k_x)$, where the coefficients
$\beta_{BIA}$ and $\beta_{SIA}$ represent the bulk- and
structure-inversion asymmetry. Applying the same reasons as in
\cite{Nature02} one can estimate the spin-galvanic contribution to
the magneto-gyrotropic photocurrent as
\begin{equation}
j \sim e \tau_p \frac{\beta}{\hbar} \frac{g \mu_B B_x}{\bar{E}}
\frac{\eta I}{\hbar \omega}\: \xi\:.
\end{equation}
Here $\xi$ is a factor describing the depolarization of electron
spin under photoexcitation followed by energy relaxation. It can
be estimated as $\xi \sim \tau_{\varepsilon}/\tau_{s}$, where
$\tau_{\varepsilon}$ is the time of electron energy relaxation
governed mainly by electron-electron collisions, and $\tau_{s}$ is
the spin relaxation time. Assuming the times to be
$\tau_{\varepsilon} \sim 10^{-13}\,s$ and $\tau_{s} \sim
10^{-10}\,s$ for the room temperature, the factor $\xi$ can be
estimated as $10^{-3}$.

Finally, the diamagnetic mechanism is related to spin-independent
magnetic field induced ${\bm k}$-linear terms, $\gamma_{\nu} (B_x
k_y - B_y k_x)$, in the electron Hamiltonian, where $\gamma_{\nu}
= ( e \hbar /c m^*) \bar{z}_{\nu}$ and $\bar{z}_{\nu} = \langle e
\nu \vert z \vert e \nu \rangle$ is the center of mass of the
electron envelope function in the $\nu$-th subband
\cite{andostern}. In an asymmetrical QW, $\bar{z}_{\nu}$ are
nonzero and the subband dispersion is given by parabolas with
their minima (or maxima in case of the valence band) shifted from
the origin $k_x=k_y=0$ by values proportional to the in-plane
magnetic field. One can show that the diamagnetic mechanism
contributes to the photocurrent under indirect intra-subband
optical transitions only if the indirect transition  involves
intermediate states in other bands or subbands, $n \neq e1$. Then,
the magneto-gyrotropic photocurrent is estimated as
\begin{equation}
j_y \sim e \tau_p \frac{\gamma B_x}{\hbar} \frac{\eta I}{\hbar
\omega}\: \xi'\:, \label{3}
\end{equation}
where $\gamma$ is a linear combination of the constants
$\gamma_{e1}$ and $\gamma_{\nu}$ in the subband $e1$ and
intermediate band. For intermediate states in the valence band,
the parameter $\xi'$ is of the order of $(\hbar \omega /E_g)^3$,
and, for intermediate states in the next conduction subband $e2$,
one has $\xi' \sim (\hbar \omega / E_{21})^3$, where $E_g$ is the
band gap, and $E_{21}$ is the $e2$-$e1$ energy spacing.

It is reasonable to expect the ratios $R_{xx}/P$ and $R_{xy}/P$
 to be of order of
$\beta_{BIA}/\bar{E}$ and $\beta_{SIA}/\bar{E}$, respectively. The
ratio $\beta_{SIA}/\gamma$ can be obtained by using the equation
for $\beta_{SIA}$ derived in Ref.~\cite{pfeffer99}. It follows
then that, because of the small parameters $\xi$ and $\xi'$, it is
the first mechanism that is responsible for the magneto-gyrotropic
photocurrent induced in InAs/AlGaAs QWs under indirect absorption
of the far-infrared radiation.


\section{Summary}

In conclusion, it has been shown that in gyrotropic systems  the
application of an external magnetic field induces in addition to
the helicity-sensitive spin-galvanic current a light polarization
independent current which we attribute to the magneto-gyrotropic
effect. Three different microscopic mechanisms that may contribute
to this effect have been considered. Under steady-state optical
excitation, the role of each mechanism in the magneto-gyrotropic
photocurrent is established through theoretical estimation.
However, the three contributions can be separated experimentally
in time-resolved measurements. Indeed, after the removal of light
or under ultra-short pulsed photoexcitation the magneto-gyrotropic
current should decay, for the mechanisms considered above, within
the energy ($\tau_{\varepsilon}$), spin ($\tau_s$) and momentum
($\tau_p$) relaxation times, respectively. A photogalvanic
detection of the nuclear spin resonance could be another evidence
for spin-related mechanisms of the photocurrent.

\section{Acknowledgements}
The high quality quantum wells were kindly provided by J.~De~Boeck
and G.~Borghs from IMEC Belgium.  We thank L.E.~Golub for helpful
discussion.

\end{document}